\documentclass{elsart}
\usepackage{natbib}
\usepackage{epsfig}
\begin{document}
\runauthor{Ballesta and Manneville}
\begin{frontmatter}
\title{The Faraday instability in wormlike micelle solutions}
\author[CRPP]{P. Ballesta}
\author[CRPP,ENSL]{S. Manneville}

\address[CRPP]{Centre de Recherche Paul Pascal -- CNRS UPR8641\\Avenue Schweitzer, 33600 PESSAC, France}
\address[ENSL]{Present address: Laboratoire de Physique ENS Lyon -- CNRS UMR5672\\46 all\'ee d'Italie, 69364 LYON cedex 07, France}

\begin{abstract}
The behaviour of semi-dilute wormlike micelle solutions under vertical vibrations is investigated using classical measurements of the Faraday instability threshold along with two new experiments based on particle imaging velocimetry and birefringence. We provide evidence for the presence of oscillations of the critical acceleration and wave number with the vibration frequency, which are linked to oscillations of the velocity and birefringence fields along the vertical direction. These observations are interpreted in terms of standing elastic waves across the fluid layer. Such an interpretation is confirmed numerically by using the model proposed by S. Kumar [Phys. Rev. E, {\bf 65}, 026305 (2002)] for a viscoelastic fluid to calculate the velocity and deformation fields. Finally, further birefringence experiments above the instability threshold show that the Faraday instability does {\it not} induce disentanglement and breakage of the micellar network.
\end{abstract}
\begin{keyword}
Wormlike micelles, Faraday instability, particle imaging velocimetry, birefringence 
\end{keyword}
\end{frontmatter}

\section*{Introduction}

In the classical ``Faraday experiment'' a fluid layer is submitted to vertical vibrations at a frequency $f$ and acceleration $a$~\cite{Faraday:1831}. By increasing the acceleration from zero, the system undergoes a bifurcation from a simple, well-defined state (flat interface with zero velocity in the bulk) to a more complex, destabilized state: at a given acceleration $a_c$ called the critical acceleration, the flat interface goes unstable and a surface wave pattern characterized by the critical wave number $k_c$ develops~\cite{Faraday:1831,Rayleigh:1883,Benjamin:1954,Chen:1997}. When the driving acceleration is further increased above $a_c$,  secondary instabilities take place~\cite{Edwards:1994}, leading at even higher accelerations to spatio-temporal chaos \cite{Kudrolli:1996}. Together with the Rayleigh-B\'enard convective instability, the Faraday instability has often been described as a paradigm for the study of dynamical systems at the laboratory scale \cite{Manneville:1990}.

Until about a decade ago, the Faraday instability had been intensively studied in {\it Newtonian} fluids~\cite{Chen:1997,Edwards:1994,Kudrolli:1996,Kumar:1994,Wernet:2001}. Only recently have vertically vibrated complex fluids been the subject of experimental~\cite{Raynal:1999,Wagner:1999,Merkt:2004,Huber:2005,Kityk:2006} and theoretical research~\cite{Kumar:1999,Muller:1999}. It was shown that the coupling between the microstructure of complex fluids and the Faraday instability either only slightly modifies the instability (e.g. by affecting the critical acceleration \cite{Raynal:1999,Wagner:1999} or by delaying the transition to disordered states \cite{Kityk:2006}) or presents non-intuitive behaviours such as stabilized ``persistent holes'' in shear-thickening fluids~\cite{Merkt:2004}. With the notable exception of concentrated colloidal and granular suspensions \cite{Merkt:2004,Lioubashevski:1999}, most of these works were performed on fluids whose viscoelasticity is only a small perturbation~\cite{Raynal:1999,Huber:2005,Kityk:2006} to Newtonian behaviour. The case of a complex fluid with significant viscoelasticity was addressed in Ref.~\cite{Wagner:1999} and it was shown that viscoelasticity could lead to a harmonic response of the fluid surface ({\it i.e.} at $f$) instead of the classical subharmonic response ({\it i.e.} at $f/2$).

In a previous paper~\cite{Ballesta:2005} we reported onset measurements of the Faraday instability in a semi-dilute surfactant solution (CPyCl--NaSal at $4\%$~wt.) known to form an entangled network of ``wormlike'' micelles. In this very viscoelastic fluid, the critical acceleration and wave number were shown to present {\it oscillations} as a function of the driving frequency $f$. We interpreted these oscillations in terms of {\it standing elastic waves} between the disturbed surface and the container bottom.

The present article is devoted to a more thorough study of this striking effect of viscoelasticity on the Faraday instability. After a brief description of our wormlike micellar system and of our experimental setup, we first present new sets of Faraday experiments in various wormlike micelle solutions and explore the dependence of the fluid response on surfactant concentration and on temperature. Then, the interpretation in terms of standing elastic waves is confirmed using two novel experiments: particle imaging velocimetry (PIV) and birefringence under vertical vibrations, which show a non monotonic behaviour of the velocity and deformation fields along the vertical direction. We also use the finite-depth model of ref.~\cite{Kumar:1999} to predict the velocity field. This numerical calculation accounts qualitatively for all our experimental observations. Finally, reasons for quantitative discrepancies between the model and the experiments are discussed. In particular, birefringence measurements above onset show that the specific nonlinear features of the rheology of wormlike micelles, namely disentanglement and breakage of the micellar network and shear-induced alignment, do not come into play in our experiments.

\section{Viscoelastic fluid: wormlike micelle solution}

Wormlike micelles are long (typically hundreds of nanometers to tens of microns), cylindrical (a few nanometers in diameter) and flexible aggregates of surfactant molecules in aqueous solution \cite{Larson:1999}. Depending on the concentration and on the surfactant, wormlike micelle solutions present various characteristic behaviours: shear-thickenning at low concentration~\cite{Liu:1996}, shear-induced alignment in the semi-dilute regime~\cite{Berret:1997}, and equilibrium isotropic to nematic phase transition at even larger concentrations~\cite{Berret:1994}. In the following we focus on {\it semi-dilute} solutions~\cite{Berret:1993}. As shown by Cates~\cite{Cates:1987}, the unique viscoelastic behaviour, namely almost a perfect Maxwell fluid \cite{Rehage:1988}, of these solutions is due to the interplay of two processes: reptation motion as in conventional polymers~\cite{Bird:1987} and breaking and recombination of the micelles under thermal agitation. Therefore, the complex viscosity of these solutions can be written:
\begin{equation}
\label{eqO} \eta^* (\omega ) =\eta_0 \frac{1+i\omega \tau_2}{1+i\omega \tau_1},
\end{equation}
where $\eta_0$ is the fluid zero-shear viscosity, $\tau_1$ the relaxation time, $\omega$ the pulsation, and $\tau_2$ a second characteristic time.

For a pure Maxwell fluid, $\tau_2=0$. However, due to the presence of the solvent, we shall rather use $\tau_2=\tau_1\eta_S/\eta_0$, where $\eta_S\ll\eta_0$ is the solvent viscosity (in our experiments, $\eta_S=\eta_{brine}=10^{-3}$~Pa.s and $\eta_0\simeq 10$~Pa.s). Thus, in the case of a Maxwell fluid, there are really only two free parameters $\eta_0$ and $\tau_1$ in eq.~(\ref{eqO}) since $\tau_2$ simply derives from the solvent viscosity $\eta_S$. Another widely used model for the linear rheology of wormlike micelles is the Oldroyd fluid whose complex viscosity is also given by eq.~(\ref{eqO}) but where $\tau_2$ is left as a free parameter (see {\it e.g.} refs.~\cite{Manero:2002,Yesilata:2006}). We shall see below that this additional free parameter allows one to account for the behaviour of our solutions over a much larger frequency range than the Maxwell fluid. Note that such a use of the Oldroyd model assumes the validity of the Cox-Merz rule which may be violated in some wormlike micelle solutions for $\omega \tau_1> 1$ \cite{Manero:2002,Kadoma:1997}. We still use fits by the Oldroyd model since equation~(\ref{eqO}) yields good empirical representation of our data. Also note that more elaborate models that account not only for the solvent but also for high frequency relaxation modes based on Rouse or Zimm approaches may apply to wormlike micelle solutions \cite{Fischer:1997}. Finally we introduce the shear modulus as the high-frequency limit of the imaginary part of $-\omega \eta^* (\omega)$ {\it i.e.} $G_0=\lim_{\omega\rightarrow \infty } \Im (-\omega \eta^* (\omega))=\eta_0 (\tau_1-\tau_2)/\tau_1^2\simeq \eta_0/\tau_1$.

In the following, we study various wormlike micelle solutions made of cetylpyridinium chloride (CPyCl, from Aldrich) and sodium salicylate (NaSal, from Acros Organics) dissolved in brine (0.5~M NaCl) with a fixed concentration ratio [NaSal]/[CPyCl]=0.5 as described in refs.~\cite{Berret:1997,Rehage:1988}. The CPyCl--NaSal concentration of our solutions varies from $2\%$ to $8\%$~wt. This concentration range ensures that (i) the micelles are in the semi-dilute regime, {\it i.e.} remain entangled and form a viscoelastic network ~\cite{Berret:1997} and (ii) the solutions are not too viscous ($\eta_0\simeq G_0\tau_1<100$~Pa.s) so that the onset of Faraday waves can be observed.

\begin{table}   
    \begin{center}
        \begin{tabular}{|c|c|c|c|c|c|}
        \hline 
        $c$ ($\%$ wt.)      & 2         & 3         & 4         & 6         & 8         \\ \hline
        $\eta_0$ (Pa.s) &2.97       &12.8   &   27.6    &   74.1    &129    \\ \hline
        $\tau_1$ (s)            &   0.421   &0.713  &0.882  &1.08   &1.11   \\ \hline
        \end{tabular}
    \end{center}
    \caption{\label{tab0} Best fit parameters for a Maxwell fluid (eq.~(\ref{eqO}) where $\tau_2=\tau_1\eta_S/\eta_0$ is fixed) for CPyCl--NaSal solutions of various concentrations $c$.}
\end{table}

\begin{table}   
    \begin{center}
        \begin{tabular}{|c|c|c|c|c|c|}
        \hline 
        $c$ ($\%$ wt.)      &2          &3          &4          &6          &8          \\ \hline
        $\eta_0$ (Pa.s) &2.95       &12.6   &   27.7    &   77.5    &129    \\ \hline
        $\tau_1$ (s)            &   0.411     &0.714  &0.891  &1.13   &1.10   \\ \hline
        $\tau_2$ (ms)       &   2.00           &0.754        &0.611  &0.344        &0.849  \\ \hline
        \end{tabular}
    \end{center}
    \caption{\label{tab1} Best fit parameters for an Oldroyd fluid (eq.~(\ref{eqO}) where both $\tau_1$ and $\tau_2$ are free parameters) for CPyCl--NaSal solutions of various concentrations $c$.}
\end{table}

\begin{figure}
\begin{center}
\includegraphics{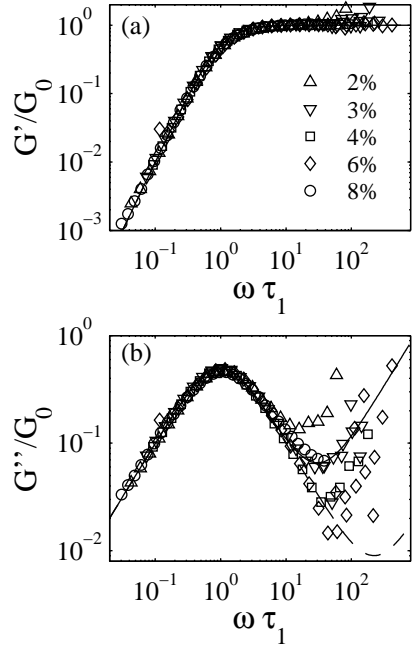}
\end{center}
\caption{Linear rheology of 2, 3, 4, 6, and $8\%$ wt. CPyCl--NaSal solutions. (a) Storage modulus $G'/G_0$ and (b) loss modulus $G''/G_0$ versus $\omega\tau_1$, where $\omega$ is the shear oscillation pulsation, $\tau_1$ the main relaxation time, and $G_0=\lim_{\omega\rightarrow \infty }G'(\omega)$ the shear modulus. The dashed line and the continuous line respectively represent a Maxwell fluid ($\eta_0=49$~Pa.s, $\tau_1=0.84$~s) and an Oldroyd fluid ($\eta_0=50$~Pa.s, $\tau_1=0.85$~s, $\tau_2=0.91$~ms). }
\label{fig0}
\end{figure}

Linear rheological measurements were performed in the cone-and-plate geometry using a  shear rate controlled rheometer (ARES, TA Instruments) with small shear rate oscillations of amplitude 0.1~s$^{-1}$ (the linear regime extends up to at least 1~s$^{-1}$ for all concentrations under study). Such measurements yield reliable values for the viscoelastic moduli $G'(\omega)=-\omega\Im [\eta^* (\omega)]$ and $G''(\omega)=\omega\Re [\eta^* (\omega)]$ as long as the oscillation frequency $f=\omega/2\pi$ remains smaller than about 40~Hz. Figure~\ref{fig0} shows the dimensionless data $G'/G_0$ and $G''/G_0$ as a function of $\omega\tau_1$ for the various solutions used in the present work. The fact that the $G''/G_0$ vs $\omega\tau_1$ data do not collapse on a single curve at high frequency provides a clear indication that the Maxwell model fails for $\omega\tau_1>3$. Indeed both $G'$ and $G''$ are well accounted for by eq.~(\ref{eqO}) over the whole frequency range provided $\tau_2$ is left as a free parameter (Oldroyd fluid). Tables~\ref{tab0} and \ref{tab1} gather the best fit parameters using eq.~(\ref{eqO}) as a function of surfactant concentration. As shown by the dashed lines in Fig.~\ref{fig0} (see also Fig.~\ref{fig4}(a) for raw data measured on a $3\%$~wt. CPyCl--NaSal solution and the corresponding Maxwell and Oldroyd fits), the Maxwell model (where $\tau_2=\tau_1\eta_S/\eta_0$ is fixed) leads to a good description of the $G''/G_0$ data only for $\omega\tau_1$ smaller than about 3. This means that higher frequency modes, which are captured by the Oldroyd model, come into play for $\omega\tau_1>3$, a feature that is classically observed in wormlike micelles \cite{Fischer:1997}. Finally a power-law fit of our $\eta_0$ vs $c$ data yields an exponent of 2.7$\pm 0.3$ in rough agreement with the value of 3.3 reported in refs.~\cite{Berret:1993,Berret:1994} and consistently with the ``fast-breaking'' limit for micellar dynamics \cite{Cates:1987}

\section{Critical acceleration and wave number measurements}

\subsection{Experimental setup}

Our experimental setup is similar to the one used in refs.~\cite{Edwards:1994,Ballesta:2005}. It consists of a cylindrical Plexiglas container of diameter $d=60$~mm filled to a height $h=10$~mm under the brimful boundary condition \cite{Edwards:1994}. This cell is sealed by a Plexiglas cover that prevents evaporation and surface contamination, and vertically vibrated by an electromagnetic shaker (Ling Dynamic Systems V406). Except when stated otherwise, the fluid temperature is controlled to $21\pm 0.5^{\circ}$C by water circulation beneath the container. The acceleration is measured by a small piezoelectric accelerometer (Endevco 2224C) attached to the cell. The wave number is inferred from images of the surface taken by a CCD camera (Cohu). The illumination technique allows us to detect surface deformation down to an amplitude of about $30~\mu$m \cite{BallestaPhD:2006}. We checked that our container is large enough for the system to be considered as laterally unbounded by performing experiments in various Newtonian fluids (silicon oils and water--glycerol mixtures) and by comparing the experimental measurements of $a_c$ and $k_c$ to the numerical calculation of ref.~\cite{Kumar:1994} for an unbounded viscous fluid. As already mentioned in ref.~\cite{Ballesta:2005}, the agreement between experiment and theory is almost perfect over the whole range of frequencies $f=20$--120~Hz, so that finite-size effects due to lateral boundaries are negligible.

\subsection{Measurements of $a_c$ and $k_c$}

At a given driving frequency $f$, the instability threshold is determined as follows: (i) the acceleration is quickly increased until the surface is fully destabilized then quickly decreased until the instability completely disappears; (ii) from this last value, the acceleration is slowly increased again (by approx. $1\%$/min) until the interface goes unstable, which yields an ``upper limit'' ${a_c}_{max}$ for the critical acceleration; (iii) once the instability appears, we wait for the whole surface to be fully destabilized (which takes about 1~min); (iv) a picture of the surface is taken from which the critical wave number $k_c$ is estimated; (v) the acceleration is finally slowly decreased (by approx. $1\%$/min) until the instability completely disappears, which yields a ``lower limit'' ${a_c}_{min}$ for the critical acceleration. In all cases, the upper and lower limits of $a_c$ differ by less than $2\%$: no significant hysteresis is observed and we define $a_c$ as the average $a_c=({a_c}_{min}+{a_c}_{max})/2$. For all our solutions and over the whole range of investigated frequencies $f=20$--120~Hz, the surface response was found to be {\it subharmonic} {\it i.e.} the standing wave pattern oscillates at $f/2$.

\begin{figure}
\begin{center}
\includegraphics*{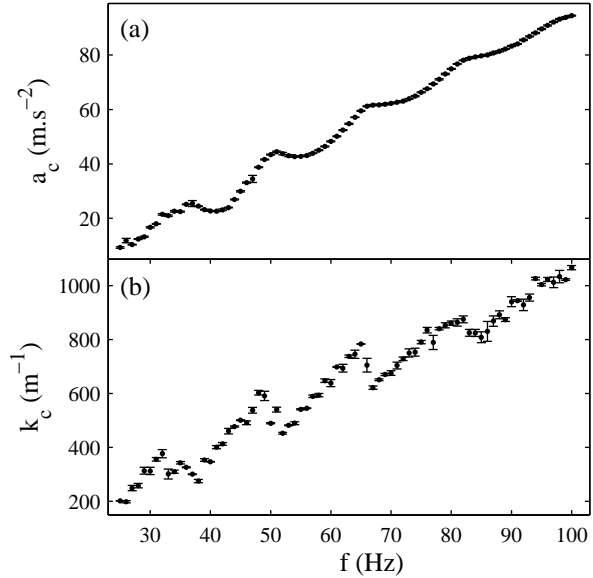}
\end{center}
\caption{(a) Critical acceleration $a_c$ and (b) critical wave number $k_c$ plotted against the vibration frequency $f$ for a $3\%$~wt. CPyCl--NaSal solution at $T=21^{\circ}$C. Error bars on $a_c$ are estimated from the experimental ``upper'' and ``lower'' limits for $a_c$ by $\pm ({a_c}_{max}-{a_c}_{min})/2$ (see text). Error bars on $k_c$ correspond to the standard deviation of 3 to 6 estimations performed on various areas of the picture of the surface.}
\label{fig1}
\end{figure}

As reported in ref.~\cite{Ballesta:2005}, the critical acceleration and wave number in wormlike micelles strikingly differ from their equivalent in Newtonian fluids. Indeed both curves $a_c$ and $k_c$ present pronounced oscillations as a function of the driving frequency $f$. Figure~\ref{fig1} shows $a_c$ and $k_c$ vs $f$ for a $3\%$~wt. CPyCl--NaSal solution (see also fig.~1 of ref.~\cite{Ballesta:2005} for similar measurements on a $4\%$~wt. solution). A consequence of this peculiar behaviour is that $k_c$ does not follow the usual dispersion relation for gravito-capillary surface waves: $\omega^2 =(gk_c+\sigma k_c^3/\rho)\tanh(k_c h)$, where $\omega=\pi f$ for subharmonic response, $g$ is the acceleration due to gravity, $\sigma$ the surface tension and $\rho$ the fluid density~\cite{Benjamin:1954}. Instead we find a non-monotone dispersion relation. Since the same experiments performed on Newtonian fluids with similar viscosity and surface tension do not display such behaviour, the  oscillations observed in fig.~\ref{fig1} must be linked to the viscoelastic properties of our micellar solutions.

\subsection{Simple physical interpretation of the oscillations}

In ref.~\cite{Ballesta:2005} we proposed to interpret the oscillations in $a_c$ and $k_c$ as the signature of ``standing elastic waves'' across the cell. Indeed in a strongly viscoelastic fluid, the disturbed surface is likely to generate shear waves that propagate downward across the fluid layer, are reflected at the bottom of the cell, backpropagate toward the surface where they may interact with the surface wave provided the attenuation over a distance $2h$ is not too large. The velocity of shear waves is given by $c(\omega)=\sqrt{G'(\omega)/\rho}\simeq\sqrt{G_0/\rho}$ for a Maxwell or an Oldroyd fluid in the $\omega\tau_1\gg 1$ limit, which is always verified in our experiments (from table~\ref{tab1}, $\omega\tau_1=2\pi f\tau_1$ is in the range 25--400). Using this simplification one can calculate the phase of the reflected shear wave as it reaches the surface. Indeed, the total travel time from the surface to the bottom of the cell and back to the surface is $2h/c$, which corresponds to a phase shift $(2h/c)(\omega/2)$ since the response of the surface is subharmonic. Moreover the no slip boundary condition at the bottom of the cell leads to an additional phase shift of $\pi$ so that the phase of the reflected shear wave at the surface is $\varphi=\omega h \sqrt{\rho/G_0}+\pi$. If $\varphi=2\pi n$, with $n$ an integer, the reflected shear wave and the surface wave are in phase leading to constructive interference, amplification of the surface wave, and a local minimum of $a_c$. On the other hand, when $\varphi=(2n+1)\pi$, the interference is destructive and surface waves are hindered, leading to a local maximum of $a_c$, hence the oscillations of Fig.~\ref{fig1}(a).

This simple interpretation, which only focuses on a single reflection and neglects attenuation of the shear wave, leads to the following expression for the distance $\delta f$ between two maxima of $a_c$: 
\begin{equation}
\label{eq2} \delta f_{th} = \frac{1}{h} \sqrt{\frac{G_0}{\rho}}.
\end{equation}

\begin{figure}
\begin{center}
\includegraphics*{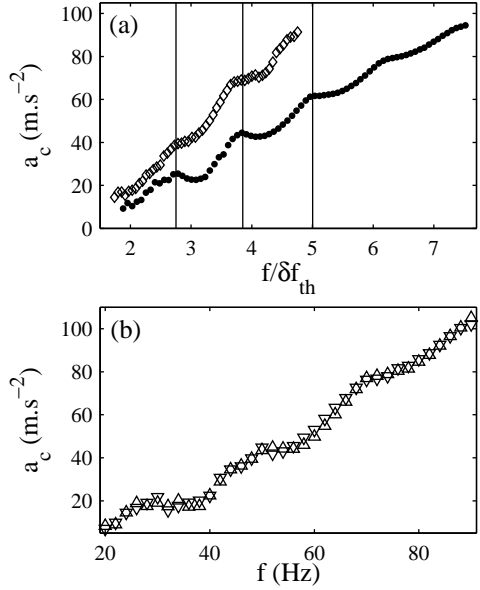}
\end{center}
\caption{(a) Critical acceleration $a_c$ versus $f/\delta f_{th}$ with $\delta f_{th}=\sqrt{G_0/\rho}/h$ for a $3\%$~wt. ($\bullet$) and a $4\%$~wt. ($\diamond$) CPyCl--NaSal solution at $T=21^{\circ}$C. (b) Critical acceleration $a_c$ versus $f$ for a $4\%$~wt. CPyCl--NaSal solution at $23^{\circ}$C ($\bigtriangleup$) and $19^{\circ}$C ($\bigtriangledown$).}
\label{fig2}
\end{figure}

The experiments presented in ref.~\cite{Ballesta:2005} were restricted to only one sample at a single surfactant concentration of $4\%$~wt. The main objective of the present work is to further check the prediction of equation~(\ref{eq2}) over the whole semi-dilute regime. More experiments were thus performed by varying the surfactant concentration, {\it i.e.} for different elastic moduli $G_0$. Figure~\ref{fig2}(a) shows the critical acceleration versus the reduced frequency $f/\delta f_{th}$ for $3\%$ and $4\%$~wt. CPyCl--NaSal solutions, where the value of $G_0$ in eq.~(\ref{eq2}) is given by linear rheological measurements. In this representation, critical accelerations oscillate roughly with the same period as a function of reduced frequency. However the maxima of $a_c$ do not exactly correspond to integer multiples of $f/\delta f_{th}$, which points to a first limitation of the simple physical picture proposed above. We shall come back to this discrepancy in sect.~\ref{snumvel}.

Equation~(\ref{eq2}) also predicts that $\delta f$ does not depend on the fluid relaxation time $\tau_1$. For CPyCl--NaSal wormlike micelle solutions, it is well known that $\tau_1$ greatly depends on the temperature while $G_0$ is nearly temperature independent~\cite{Fischer:1997}. Indeed on our $4\%$~wt. micellar solution, linear rheological experiments gave $G_0=33$~Pa and $\tau_1=0.74$~s at $T=19^{\circ}$C (smaller temperatures led to crystallization of the sample), and $G_0=32$~Pa and $\tau_1=0.37$~s at $T=23^{\circ}$C: a $4^{\circ}$C change in temperature only induces a 3\% change in $G_0$ while $\tau_1$ is doubled. Figure~\ref{fig2}(b) shows $a_c$ measurements performed at $T=19^{\circ}$C and $T=23^{\circ}$C (slightly below room temperature to prevent condensation on the Plexiglas cover). As expected from eq.~(\ref{eq2}) the experimental curves are almost undistinguishable.

\begin{figure}
\begin{center}
\includegraphics*{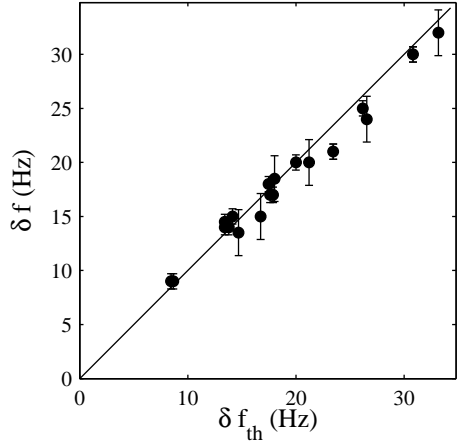}
\end{center}
\caption{Experimental $\delta f$  ($\bullet$) versus predicted $\delta f_{th}=\sqrt{G_0/\rho}/h$. The continuous line is the first diagonal.}
\label{fig3}
\end{figure}

Finally, more experiments at different fluid heights, surfactant concentrations, and temperatures were performed. Figure~\ref{fig3} gathers experimental values of $\delta f$ inferred from these measurements as a function of $\delta f_{th}$ predicted by eq.~(\ref{eq2}) where $G_0$ was extracted from linear rheology. This plot clearly shows that eq.~(\ref{eq2}) holds for all our experiments so that our interpretation in terms of standing elastic waves is relevant.

\section{Analytical and numerical predictions of the linear stability analysis}

In this section, we compare our experimental results to the numerical approach first proposed by L.~Tuckermann and K.~Kumar~\cite{Kumar:1994} and adapted to viscoelastic fluids by S.~Kumar~\cite{Kumar:1999} by taking into account a complex, frequency dependent viscosity $\eta^*(\omega)$.

\subsection{Numerical predictions for $a_c$ and $k_c$}
\label{sasl}

The numerical scheme relies on a linear stability analysis based on Floquet theory. Fourier expansions of the velocity field $\mathbf{v}$ and of the surface deformation $\xi$ together with linearization of the Navier-Stokes equations lead to a $N\times N$ eigenvalue problem, where $N$ corresponds to the cut-off in the Fourier expansions (we chose to use $N=8$, but $N=4$ already provides good estimates). The eigenvalues $a(\omega_0,k)$ correspond to marginal values of the acceleration {\it i.e.} for a given pulsation $\omega_0=2\pi f$ of the excitation, surface modes with wave number $k$ go unstable when $a>a(\omega_0,k)$. For a fixed $\omega_0$, $a=a(\omega_0,k)$ defines a set of so-called ``resonance tongues'' in the $(a,k)$ plane. Each resonance tongue is associated to a different resonance frequency $\omega=n\omega_0/2$ of the surface, where $n$ is a positive integer. Odd (resp. even) values of $n$ correspond to a subharmonic (resp. harmonic) response.

\begin{figure}[h]
\begin{center}
\includegraphics*{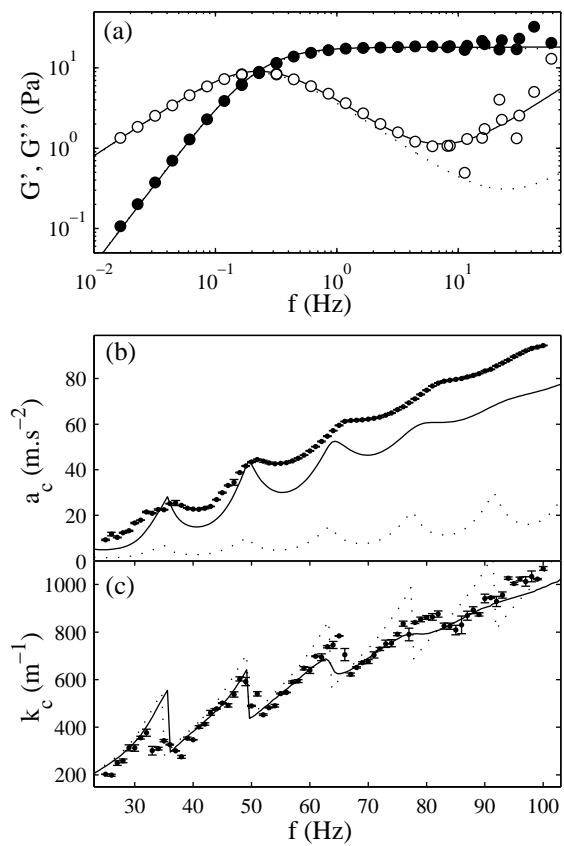}
\end{center}
\caption{(a) Linear rheology of a $3\%$~wt. CPyCl--NaSal solution: experimental viscoelastic moduli $G'$ ($\bullet$) and $G''$ ($\circ$) versus shear oscillation frequency $f$. The dotted line and the continuous line respectively represent the best fits by a Maxwell fluid over $f=0$--3~Hz (see table~\ref{tab0}), and by an Oldroyd fluid over $f=0$--40~Hz (see table~\ref{tab1}). (b) Critical acceleration $a_c$ and (c) critical wave number $k_c$ versus vibration frequency for a $3\%$~wt. CPyCl--NaSal solution. The dotted (resp. continuous) lines are the result of the numerical calculation for the previous Maxwell (resp. Oldroyd) fluid.}
\label{fig4}
\end{figure}

This linear stability analysis allows one to determine the critical acceleration $a_c$ as the global minimum of the resonance tongues and to define the corresponding wave number as the critical wave number $k_c$. In all the cases investigated in the  work, $a_c$ and $k_c$ were found to lie on the first resonance tongue, so that the surface response is always subharmonic at $\omega=\omega_0/2=\pi f$. The only severe constraint of the numerical method of ref.~\cite{Kumar:1999} is that it requires the knowledge of $\eta^*(\omega)$ over a wide range of frequencies, typically 0--$Nf/2\simeq 2f$ for an external forcing frequency $f$. Since $f$ can reach 120~Hz in our experiments, rheological measurements of $\eta^*(\omega)$ are in principle required up to about 250~Hz, which is far above the upper limit of our rheometer (about $f=40$~Hz). Thus, above $f\simeq 20$~Hz, extrapolations of $\eta^*(\omega)$ were used in the numerical calculation based on the two models discussed above: the Maxwell and the Oldroyd models.

Figure~\ref{fig4}(a) shows the best fit of the viscoelastic moduli of our $3\%$~wt. CPyCl--NaSal solution by a Maxwell fluid, which yields an accurate description of the experimental $G'$ and $G''$ up to $f\simeq 2$~Hz. Feeding the numerical calculation of ref.~\cite{Kumar:1999} by this Maxwell model leads to the values of $a_c$ and $k_c$ shown as dotted lines in fig.~\ref{fig4}(b) and (c). The first noticeable result is the presence of strong oscillations in $a_c$ with the same periodicity as the experimental ones. Moreover the evolution of $k_c$ is qualitatively similar to that of the experimental data. Still experiment and theory remain quantitatively different. The critical acceleration predicted numerically is smaller than the experimental $a_c$ by about an order of magnitude and, for both $a_c$ and $k_c$, the amplitude of the oscillations predicted from the Maxwell model increases with frequency whereas it is clearly seen to decrease in the experiment. Since the vibration frequency is always greater than $20$~Hz, these discrepancies are most probably linked to the poor fit of $G''$ by the Maxwell model for $f>2$~Hz.

To account for viscoelasticity at higher frequencies, the rheological measurements were fit by an Oldroyd fluid. As already pointed out, the Oldroyd model has two free parameters $\tau_1$ and $\tau_2$ and leads to a better fit of the viscoelastic moduli for $f>2$~Hz (see the continuous lines in fig.~\ref{fig4}(a)). The linear stability analysis using the corresponding Oldroyd fluid also yields a much better agreement between numerical and experimental $a_c$ and $k_c$. In particular the amplitude of the oscillations now decreases with increasing frequency as seen in the experiment. However, for $f>60$~Hz, the critical acceleration found numerically remains smaller than the experimental one by about 20\%.

At this stage the lack of rheological data at frequencies larger than 40~Hz does not allow us to draw any definite conclusion on whether a better agreement between theory and experiment would be achieved by using a more elaborate rheological model at high frequencies. Experimental imperfections such as absorption of shear waves at the bottom of the container or additional dissipation at the walls, which are not taken into account in the numerical calculation, may also account for the observed discrepancies. Another possibility is the fact that surface waves may affect the microstructure of the fluid in a nonlinear way (see sect.~\ref{sbiref} below). Therefore we did not try to look further for more appropriate rheological models but we rather focused on the original features of the Faraday instability in our strongly viscoelastic fluid, namely the presence of elastic waves across the sample.

\subsection{Calculation of the velocity field}
\label{snumvel}

Interestingly the velocity field $\mathbf{v}$ of a vertically vibrated viscoelastic fluid at onset can be inferred from the numerical model of ref.~\cite{Kumar:1999}. Indeed, if the velocity vector is expanded in terms of a Fourier series
\begin{equation}
\mathbf{v}=u\,\mathbf{e}_x+v\,\mathbf{e}_y+w\,\mathbf{e}_z=
\sum_{n=0}^{\infty}\mathbf{v}_n(x,y,z) e^{i(n +\alpha)\omega_0 t}\,,\label{eqv}
\end{equation}
with $\alpha=0$ (resp. $\alpha=1/2$) for a harmonic (resp. subharmonic) response, then ref.~\cite{Kumar:1999} shows that the $n^{\hbox{\rm\tiny th}}$ Fourier mode of the vertical component $w$ of the velocity reads
\begin{eqnarray}
w_n(x,y,z) &=& \mathbf{v}_n(x,y,z)\cdot\mathbf{e}_z = f(x,y) w_n(z)\label{eq4}\\
&=&f(x,y) \left( a_1 e^{k_c z} + a_2 e^{-k_c z} + a_3 e^{q_n z} + a_4 e^{-q_n z} \right)\,,\label{eq5}
\end{eqnarray}
where
\begin{eqnarray}
\left(\frac{\partial^2}{\partial x^2}+\frac{\partial^2}{\partial y^2}\right)f(x,y)&=&k_c^2 f(x,y)\,,\\
q_n^2&=&k_c^2+i\frac{(n+\alpha)\omega_0 \rho}{\eta^*((n+\alpha)\omega_0)}\,,\label{eqq}
\end{eqnarray}
and $k_c$ is the critical wave number.

The $a_i$'s are deduced from the boundary conditions and depend on $n$, $\alpha$, $k_c$, $\eta^*$, $\omega$, and $h$ (see ref.~\cite{Kumar:1999} for their full expressions). They are also proportional to the $(n+\alpha)^{\hbox{\rm\tiny th}}$ Fourier component of the surface wave amplitude $\xi_n$. However, since eq.~(\ref{eq5}) derives from a linear stability analysis, the velocity field is only known up to some multiplicative constant and a weakly nonlinear approach would be required to get a quantitative estimate of the velocity amplitude as a function of the distance from instability threshold. In the present work, we shall assume that our experiments are performed close enough to onset so as to allow us to use eq.~(\ref{eq5}) where the absolute value of the velocity will be left as a free parameter. Finally, the horizontal velocity $u\,\mathbf{e}_x+v\,\mathbf{e}_y$ can be calculated from $w$ and the incompressibility equation $\nabla\cdot\mathbf{v}=0$.

\begin{figure}
\begin{center}
\includegraphics*{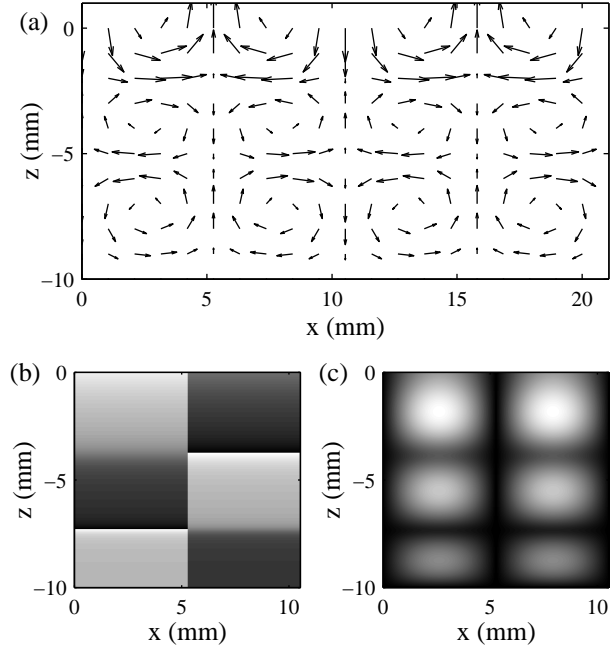}
\end{center}
\caption{(a) Numerical velocity field $\mathbf{v}(x,z,t_0=0)$ for an Oldroyd fluid with $\eta_0=20$~Pa.s, $\tau_1=0.5$~s, and $\tau_2=0.5$~ms vibrated at $f=70$~Hz. (b)  Phase $\phi(x,z)$ and (c) amplitude $u(x,z)$ of the $x$-component of the velocity $\mathbf{v}(x,z,t)$.}
\label{fig5}
\end{figure}

Figure~\ref{fig5} presents the first subharmonic mode $(n=0,\alpha=1/2)$ of the velocity field calculated from eq.~(\ref{eq5}) for an Oldroyd fluid. The velocity field consists of an array of counterrotating rolls whose number across the fluid height increases with increasing frequency. Of course, these counterrotating rolls are not steady rolls but oscillate in time at the frequency of the surface response. As shown by eq.~(\ref{eqq}), oscillations along the $z$ direction arise from the term $a_3 e^{q_n z} + a_4 e^{-q_n z}$ in eq.~(\ref{eq5}). Indeed, since $q_n$ is a complex number, it can be rewritten as $q_n=i k_{n}+1/l_{n}$, where $ k_{n}$ is the vertical wave number and $l_{n}$ is the attenuation length for the $(n+\alpha)^{\hbox{\rm\tiny th}}$ mode. 

Focusing on the case $(n=0,\alpha=1/2)$ and neglecting higher-order modes in eq.~(\ref{eq5}), it is straightforward to understand why couterrotating rolls should be detected in experiments on wormlike micelles but should remain unseen in Newtonian (or weakly viscoelastic) fluids. Indeed, with $k_c=600$~m$^{-1}$ and $\omega_0=380$~rad.s$^{-1}$ in standard experimental conditions and for an Oldroyd fluid with typical parameters $\eta_0=20$~Pa.s, $\tau_1=0.5$~s, and $\tau_2=0.5$~ms, one gets $l_{0}=16$~mm, which is greater than the vertical wavelength $2\pi/k_{0}=9$~mm. On the other hand, for a Newtonian fluid of same absolute viscosity $\eta=|\eta^*(\omega_0/2)|=0.43$~Pa.s, the calculation yields $l_{0}=1.6$~mm~$\ll 2\pi/k_{0}=36$~mm. Note also that in the latter case $l_{0}<h=10$~mm so that elastic waves are fully damped when they reach the bottom of the cell and no stationary elastic wave pattern can form across the fluid height.

Finally eq.~(\ref{eqq}) also allows one to recover eq.~(\ref{eq2}) in the case of an Oldroyd fluid and when only the first subharmonic mode is considered. In such a case, inserting eq.~(\ref{eqO}), $n=0$, and $\alpha=1/2$ in eq.~(\ref{eqq}) leads to:
\begin{equation}
q_0^2=k_c^2 +i\,\,\frac{\omega \rho}{\eta_0}\,\,\frac{1+i\omega\tau_1}{1+i\omega\tau_2}\,,
\label{eqq0}
\end{equation}
where $\omega=\omega_0/2$ is the pulsation of the surface response. Since in our experiments $\omega\tau_2\ll 1\ll\omega\tau_1$ (see table~\ref{tab1}) and $k_c^2 \eta_0/\rho \omega_0^2 \tau_1 \ll 1$, the zero-order approximation of eq.~(\ref{eqq0}) yields:
\begin{equation}
q_0 \simeq i k_0 \simeq i\omega \sqrt{\frac{\rho}{G_0}}\,.
\end{equation}
As explained above the oscillations of $a_c$ and $k_c$ are linked to the wave number $q_0$. More precisely one period of the oscillation of $a_c$ corresponds to an increase of $k_0$ by $\pi/h$ (addition of a local maximum of $\mathbf{v}$) {\it i.e.} an increase of $\omega$ by $\pi\sqrt{G_0/\rho h^2}$. For a subharmonic response this leads to $\delta f=\sqrt{G_0/\rho h^2}$, {\it i.e.} eq.~(\ref{eq2}). Thus our simple interpretation in terms of standing elastic waves is consistent with the analytical expression (\ref{eqq}) at least in a first approximation.

Expanding $q_0$ to first order, one gets :
\begin{equation}
q_0 \simeq i\omega \sqrt{\frac{\rho}{G_0}}-\frac{i k_c^2}{2\omega}\sqrt{\frac{G_0}{\rho}} + \sqrt{\frac{\rho}{G_0}}\,\frac{1+\tau_1 \tau_2 \omega^2}{2\tau_1}\,.
\end{equation}
With $q_0=i k_0+1/l_0$, the previous equation leads to
\begin{eqnarray}
k_0 &=& \omega \sqrt{\frac{\rho}{G_0}}-\frac{k_c^2}{2\omega}\sqrt{\frac{G_0}{\rho}}\label{eqk0}\,,\\
l_0 &=& \sqrt{\frac{G_0}{\rho}}\,\frac{2\tau_1}{1+\tau_1 \tau_2 \omega^2}\label{eql0}\,.
\end{eqnarray}
Since $k_c$ increases roughly linearly with $\omega$, eq.~(\ref{eq2}) becomes less accurate as frequency increases due to the corrective term in eq.~(\ref{eqk0}). This explains why maxima of $a_c$ do not correspond to integer values of $f/\delta f_{th}$ in fig.~\ref{fig2}(a). Moreover eq.~(\ref{eql0}) shows that the attenuation length $l_0$ decreases with $\omega$. At high frequencies oscillations should thus disappear. This accounts qualitatively for the decrease of the amplitude of the oscillations in $a_c$ and $k_c$ observed experimentally and numerically.

So far we have shown that the linear stability analysis of ref.~\cite{Kumar:1999} predicts an oscillating behaviour for $a_c$ and $k_c$. The calculation of the velocity field close to instability threshold unveils the existence of counterrotating rolls. Such rolls, unseen in Newtonian fluids, are a clear signature of viscoelasticity in the Faraday experiment. Our next step is to provide direct experimental evidence for the existence of these ``elastic rolls.''

\section{Experimental study of the velocity field}

In this section we first describe the experimental apparatus used to access the velocity field in the Faraday experiment. Measurements are then presented and discussed in light of the above theoretical predictions.

\subsection{Particle imaging velocimetry setup}

In order to provide experimental evidence for the existence of large-scale counterrotating rolls in our viscoelastic fluid, particle imaging velocimetry (PIV) experiments were performed under vibrations. Since wormlike micelle solutions are transparent, the scattering properties of our system were enhanced by seeding the solutions with $0.1\%$~wt. hollow glass spheres (Sphericel, Potters Industries) of mean radius $11.7$~$\mu$m and density $1.1$. Thanks to the high zero-shear viscosity of the micellar solutions, sedimentation of the spheres is negligible over the $\sim$8~hour duration of a Faraday experiment (a rough estimate yields a sedimentation velocity of about 0.03~$\mu$m.s$^{-1}$). We also checked that the rheological properties of the micellar system were not significantly affected by the addition of $0.1\%$~wt. scatterers {\it i.e.} remain very well described by an Oldroyd fluid. The best fit parameters obtained for the Oldroyd model on a $4\%$~wt. CPyCl--NaSal solution seeded with $0.1\%$~wt. scatterers are $\eta_0=24.3$~Pa.s, $\tau_1=0.87$~s, and $\tau_2=0.55$~ms, very close to those of the corresponding solution without scatterers (see table~\ref{tab1}).

In order to visualize the motion of the scatterers under vibrations, we use a parallelepipedic Plexiglas cell of length $L=180$~mm (in the $x$ direction), width $l=4$~mm (in the $y$ direction), and depth $h=10$~mm (in the $z$ direction) under the brimful boundary condition. Thanks to the large aspect ratio $L/l$, a one-dimensional surface wave pattern is forced perpendicular to the $x$ direction \cite{Douady:1990}. The cell is lit in the $(x,z)$ plane by a vertical laser sheet crossing the cell at $y_0=2$~mm. Images of the $(x,z)$ plane are recorded by a high-speed CCD camera (Mikroton MC1310) at a frequency $\sim 10f$ during 40 periods of the vertical oscillations. 

\begin{figure}
\begin{center}
\includegraphics*{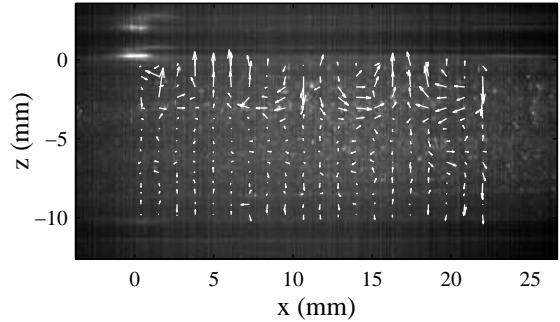}
\end{center}
\caption{Picture of a $4\%$~wt. CPyCl--NaSal solution seeded with hollow glass spheres recorded under vibrations at $f=75$~Hz by the high-speed CCD camera. The arrows show the velocity field $u\,\mathbf{e}_x+w\,\mathbf{e}_z$ computed by PIV on two consecutive images. The white spots around $x\simeq 0$ correspond to the laser beam used for detecting the bottom of the cell.}
\label{fig6}
\end{figure}

The first step in the image processing is to detect the vertical position of the cell by following a bright spot made by a laser at the bottom of the cell (around $(x\simeq 0,z\simeq -10$~mm in fig.~\ref{fig6}, see also fig.~\ref{fig12}c). The displacement field of the scatterers and the two-dimensional velocity vector $u\,\mathbf{e}_x+w\,\mathbf{e}_z$ at a given point $(x,z)$ are then computed in the moving reference frame of the vibrated cell by cross-correlating the intensity fields of two consecutive images over small square regions of size 0.25~mm$^2$. Figure~\ref{fig6} shows a typical image of the micellar solution along with the velocity field obtained by PIV under vibrations.

For the sake of clarity and since $y$ is fixed to $y_0$ in all the PIV experiments, we shall drop the dependence on $y$ of the various variables defined in the previous section whenever such dependence is not needed. Note also that the $\pm 1$~pixel uncertainty on the vertical position of the cell may remain too large compared to the vertical displacement of the fluid and in some cases leads to artifacts in the estimation of the vertical component $w$ of the velocity. Therefore, in the following, we focus on the velocity component $u$ along the $x$ direction. In order to compare experimental results to numerical calculations, we extract the mode oscillating at $\omega_0/2$ from $u(x,z,t)$ {\it i.e.} $u_0(x,z)$ as defined by eq.~(\ref{eqv}) with $\alpha=1/2$.

\subsection{Experimental velocity measurements compared to numerical predictions}

\begin{figure}
\begin{center}
\includegraphics*{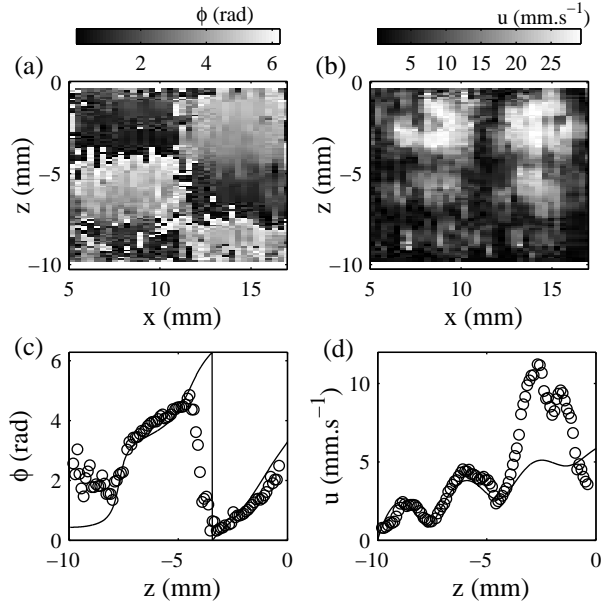}
\end{center}
\caption{(a) Phase $\phi(x,z)$ and (b) amplitude $u(x,z)$ of $u_0(x,z)$ measured by PIV in a $4\%$~wt. CPyCl--NaSal solution vibrated at $f=75$~Hz. (c) $<\phi(x,z)>$ and (d) $<u(x,z)>$ averaged over $x=6$--9~mm. The continuous lines are the predictions for $\phi$ and $u$ given by equations~(\ref{eq5}) and (\ref{equn}) with $n=0$ and $\alpha=1/2$ and using an Oldroyd fluid with $\eta_0=24.3$~Pa.s, $\tau_1=0.87$~s, and $\tau_2=0.55$~ms. Both the amplitude of $u_0$ and the reference phase are free parameters. }
\label{fig7}
\end{figure}

Shown in fig.~\ref{fig7} are the amplitude $u(x,z)$ and phase $\phi(x,z)$ of $u_0(x,z)$. It is clearly seen that the velocity does not decrease exponentially with $z$ as would be expected for a Newtonian fluid, but rather presents local maxima consistent with the presence of counterrotating rolls (see fig.~\ref{fig5}(b) and (c) for a qualitative comparison with numerical results).

In the case of our one-dimensional pattern with large aspect ratio, one may recover the horizontal velocity $u$ from the numerical calculation of $w$ by assuming $L\gg l$. Indeed under this assumption and the no slip condition at the cell border ($\mathbf{v}=0$ for $y=0,l$) one gets $v=0$ and $w(x,y,z)=\cos(k_x x)\cos(k_y y) w(z)$, with $k_y=\pi/l$ and $k_x^2=k_c^2-k_y^2$. This is equivalent to having a half mode along the $y$ direction. Next, incompressibility leads to: 
\begin{equation}
\label{equn} u_n(x,y,z)=-\frac{1}{k_x}\sin(k_x x)\cos(k_y y)\,\frac{\hbox{\rm d}w_n}{\hbox{\rm d}z}(z)\,,
\end{equation}
where $w_n(z)$ is given by eq.~(\ref{eq5}).

Figures~\ref{fig7}(c) and \ref{fig7}(d) compare the experimental results (averaged over $x=6$--9~mm) to the numerical prediction of eq.~(\ref{equn}) for $(n=0,\alpha=1/2)$ corresponding to the best fit of the linear rheological measurements by an Oldroyd fluid. The number of rolls as well as their positions in the cell are very well accounted for by the numerical calculation. We believe the discrepancies near the surface, especially the large experimental values of $u_0$, to be linked to the non-zero amplitude $\xi_0$ of the surface waves that prevents an accurate estimation of the velocity field.

\section{Birefringence experiments}
\label{sbiref}

Measurements of $a_c$ and $k_c$, experimental characterization of the velocity field close to onset, and comparison with numerical predictions have confirmed the interplay of parametrically excited surface waves and elastic waves in a strongly viscoelastic fluid under vertical vibrations. Thus the presence of a microstructure, namely an entangled network of wormlike micelles, deeply affects the Faraday instability. We now address the question raised in ref.~\cite{Ballesta:2005} of whether the instability modifies the microstructure. In particular, in the nonlinear regime, semi-dilute solutions of wormlike micelles are known to disentangle and align under shear leading to a shear-induced isotropic--nematic transition \cite{Larson:1999,Rehage:1991,Spenley:1993}. Therefore one may wonder whether the flow induced by Faraday waves is strong enough to induce such a transition, which could provide an alternative explanation for the discrepancy between experiments and linear stability analysis observed in sect.~\ref{sasl}.

Flow-induced alignment have already been much studied in semi-dilute wormlike micelle solutions using birefringence \cite{Hu:1993,Cappelaere:1997,Lerouge:2000,Decruppe:2003,Lerouge:2004,Chen:2004,Schubert:2004}. Following these previous works and in order to investigate the deformation of the microstructure under vertical vibrations, we performed birefringence experiments close to onset and further above the instability threshold. The experimental setup is modified as follows. A parallelepipedic glass cell of length $L=72$~mm and width $l=6$~mm is filled with the micellar solution to a height $h=7.1$~mm. This cell is set between two polarizers and lit from backward by an extended white light source. The high-speed CCD camera used previously for PIV measurements captures the transmitted light at about 1000~fps. For given vibration frequency $f$ and acceleration $a$, two measurements can be performed depending on the configuration of the polarizers. When the polarizers are aligned, the contour of the surface wave is easily accessed by image processing. This yields the surface deformation $\xi(x,t)$ from which the amplitude of the surface wave $\xi_0$ may be extracted. When the polarizers are crossed, the birefringence intensity $I(x,z,t)$ is recorded and analyzed. 

\subsection{First observations and qualitative discussion}

\begin{figure}
\begin{center}
\includegraphics*{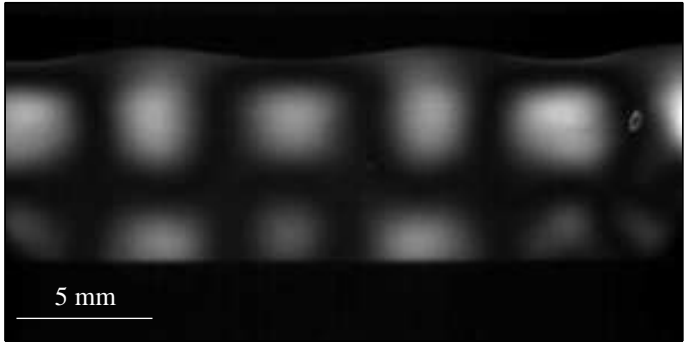}
\end{center}
\caption{Typical birefringence pattern $I(x,z,t_0)$ of an $8\%$~wt. CPyCl--NaSal solution vibrated at $f=80$~Hz.}
\label{fig8}
\end{figure}

Figure~\ref{fig8} shows a typical birefringence pattern obtained at a given time $t_0$ between crossed polarizers on an $8\%$~wt. CPyCl--NaSal solution slightly above the onset of Faraday waves. Evenly spaced bright spots fill the whole cell with the same horizontal periodicity as the surface wave. Such bright spots are observed below antinodes of the stationary surface wave. The birefringence intensity $I(x,z,t)$ oscillates in phase with $\xi(x,t)$ so that bright spots alternatively turn to black when the surface deformation goes through zero. Regions in the bulk located below nodes of the surface wave remain dark at all times.

Based on the previous analysis of the velocity field, the interpretation of the birefringence pattern is rather straightforward. Indeed, from the velocity fields of figs.~\ref{fig5} or \ref{fig6}, the two-dimensional rate of deformation tensor $\bar{\dot\gamma}$ can easily be estimated using $\bar{\dot\gamma}_{i,j}=\partial_i v_{j}$, with $i$ and $j$ being $x$, $y$, or $z$ and $(v_x,v_y,v_z)=(u,v,w)$. In the $(x,z)$ plane, $\bar{\dot\gamma}$ (not shown) is composed of localized compression and stretching zones that oscillate in time. Compression and stretching result in an oscillatory deformation of the micellar network, which leads to the observed birefringent spots.

\begin{figure}
\begin{center}
\includegraphics*{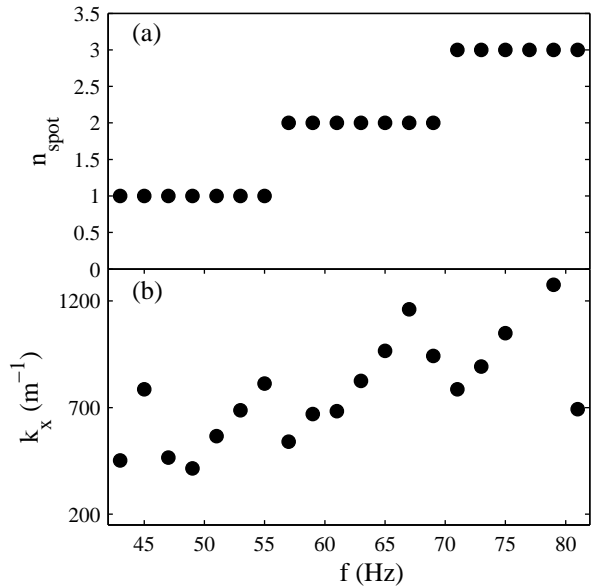}
\end{center}
\caption{(a) Number of birefringent spots $n_{spot}$ along the vertical direction and (b) wave number $k_x$ versus vibration frequency for a $5\%$~wt. CPyCl--NaSal solution.}
\label{fig10}
\end{figure}

Moreover, as shown in fig.~\ref{fig10}(a), the number $n_{spot}$ of bright spots along the vertical direction increases with increasing frequency. Figure~\ref{fig10}(b) proves that drops of $k_x$ are linked to the discretization of the number of spots along the $z$ direction. This is a direct consequence of the link between the vertical wave number and $k_c$ (see eq.~(\ref{eqq})): each time the number of spots increases by one (or equivalently each time a new set of counterrotating rolls fits in the cell height), the vertical wave number increases (by $\pi/2h$) and the surface wave number abruptly decreases.

Finally, if one defines an equivalent shear rate near the surface as $\dot\gamma=\xi_0 k_c\omega_0/2$ \cite{Kumar:1999}, a rough estimate with $\xi_0=30$~$\mu$m (the minimum amplitude that our setup can detect) yields $\dot\gamma\simeq 3$~s$^{-1}$. Since such a shear rate is of the same order of magnitude as the critical shear rate $\dot{\gamma}_c=2.6/\tau_1$ for micelle disentanglement and alignment \cite{Spenley:1993}, nonlinear rheology may come into play and we should check for the possibility of an alignment transition of the micelles under vertical vibrations. The next sections are devoted to a detailed analysis of the birefringence intensity and to more measurements above onset. 

\subsection{Birefringence measurements above onset}

For an optically anisotropic medium composed of elongated and oriented micelles, the transmitted birefringence intensity takes the following general expression \cite{Lerouge:2000}: 
\begin{equation}
I(x,z,t,\theta)=I'\sin ^2 \left( \frac{\delta(x,z,t)}{2} \right) \sin ^2 2(\chi (x,z,t)-\theta)\,,
\label{eqI}
\end{equation}
where the retardation $\delta$ reads
\begin{equation}
\delta(x,z,t) = \frac{2\pi l}{\lambda} \Delta n(x,z,t),
\end{equation}  
$I'$ is the incident intensity, $\lambda \sim 500$~nm is the light wavelength, and $\Delta n(x,z,t)$ is the birefringence intensity. $\chi(x,z,t)$ is defined as the angle between some reference direction and the average micelle direction. $\theta$ is the angle between the polarization of the incident beam and the reference direction.

To address the issue of whether or not the nonlinear deformation regime is reached and the micelles disentangle, we shall make use of the so-called ``stress-optical rule'' according to which the quantity $\Delta n \sin 2\chi$ is proportional to the shear stress $\sigma$ \cite{Lerouge:2000,Decruppe:2003}. This rule was shown to hold at low shear stresses in CTAC wormlike micelle solutions for various salt (NaSal) concentrations~\cite{Decruppe:2003}. Although to our knowledge no experimental validation of the stress-optical rule is available for our CPyCl--NaSal in brine system, we may assume that it is general enough to also hold in our experiments.

As long as the network structure remains intact and the micelles remain entangled, the deformation is elastic, {\it i.e.} $\sigma \propto \gamma$, so that one should have $\Delta n \sin 2\chi \propto \gamma$, where $\gamma$ stands for the deformation. However the stress-optical rule no longer applies when the micelles disentangle and the network is broken \cite{Lerouge:2000,Decruppe:2003}.

In the case $\delta \ll 1$ and for $\theta=0$, which is always verified for wormlike micelles \cite{Lerouge:2000}, eq.~(\ref{eqI}) and the stress-optical rule lead to
\begin{equation}
I(x,z,t) \propto \Delta n(x,z,t)^2 \sin ^2 2\chi (x,z,t)\propto \gamma(x,z,t) ^2 \,.
\label{eqI2}
\end{equation}
Focusing on the first subharmonic mode, eq.~(\ref{eqI2}) shows that $I(x,z,t)$ should oscillate at the forcing pulsation $\omega_0$ with an amplitude $I_0(x,z)\propto \gamma_0(x,z)^2$, where $\gamma_0(x,z)$ denotes the amplitude of the first subharmonic mode of $\gamma(x,z,t)$. Experimentally $\gamma_0$ is varied by increasing the driving acceleration $a$ above the critical acceleration $a_c$. Indeed the deformation at the surface $\gamma_0(x,z=\xi_0(x))$ is directly proportional to the surface wave amplitude through $\gamma_0=k_c \xi_0$. Thus increasing $a$ leads to larger surface wave amplitudes and to larger deformations, which allows one to test eq.~(\ref{eqI2}).

\begin{figure}
\begin{center}
\scalebox{1}{\includegraphics*{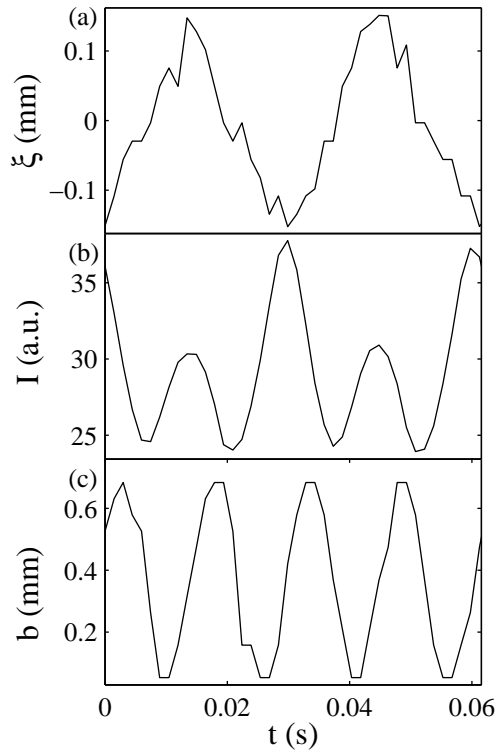}}
\end{center}
\caption{(a) Surface wave amplitude $\xi(x_0,t)$, (b) birefringence intensity $I(x_0,z_0,t)$, and (c) position of the bottom of the cell $b(t)$ for an $5\%$~wt. CPyCl--NaSal solution vibrated at $f=65$~Hz. In (b) $x_0=7.5$~mm and $z_0=-2.1$~mm were chosen to coincide with a birefringence spot. }
\label{fig12}
\end{figure}

More precisely, we first extract the position of the surface $\xi(x,t)$ from images recorded with aligned polarizers. The subharmonic amplitude of the deformation $\xi_0(x)$ is then computed and fitted to $\xi_0(x)=\xi_0|\cos(k_x x)|$  (see fig~\ref{fig12}(a)), which yields the amplitude of the deformation $\gamma_0=k_c \xi_0$ at the surface. Since the dependence of $\gamma_0(x,z)$ on $z$ is decoupled from the dependence on $x$ (see also eq.~(\ref{eq4})), one has $\gamma_0(x,z)=\gamma_0 |\cos(k_x x)| f(z)$, where $f(z)$ does not need to be specified in order to test the scaling law (\ref{eqI2}).

\begin{figure}
\begin{center}
\includegraphics*{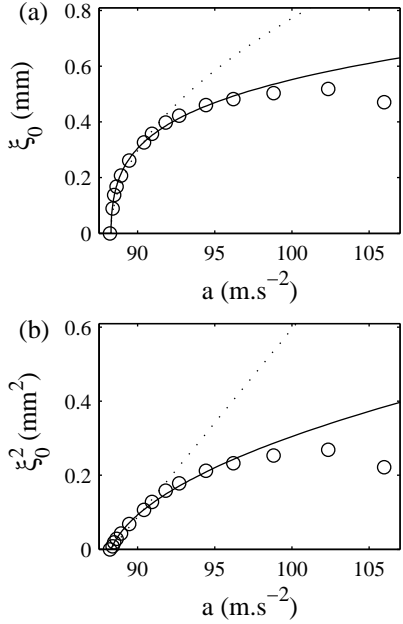}
\end{center}
\caption{(a) Surface wave amplitude $\xi_0$ versus acceleration $a$ for an $8\%$~wt. CPyCl--NaSal solution vibrated at $f=80$~Hz. (b) $\xi_0^2$ versus $a$. The dotted line is the best fit by eq.~(\ref{eq_fit0_xi}) over $a=88.3$--$90.1$~m.s$^{-2}$ and the continuous line is the best fit by eq.~(\ref{eq_fit_xi}) over $a=88.3$--$97$~m.s$^{-2}$. }

\label{fig11}
\end{figure}

Figure~\ref{fig11} shows the evolution of the surface wave amplitude $\xi_0$ with the driving acceleration $a$ above onset. In the case of the Faraday instability, it is well known that $\xi_0$ should obey the following amplitude equation \cite{Douady:1990}: 
\begin{equation}
\label{eq.diff} \tau_g \frac{d\xi_0}{dt}= \epsilon \xi_0 - \frac{ \xi_0^3}{l_3^2} - \frac{ \xi_0^5 }{l_5^4}+\,...\,,
\end{equation} 
where $\tau_g$ is the characteristic growth time of the instability, $\epsilon=a/a_c-1$ is the reduced acceleration, and $l_3$ and $l_5$ are characteristic lengths linked to the dissipation process. At steady state eq.~(\ref{eq.diff}) becomes: 
\begin{equation}
\label{eq.rest} \epsilon \xi_0 - \frac{ \xi_0^3}{l_3^2} - \frac{ \xi_0^5 }{l_5^4}+\,... = 0\,.
\end{equation}
If the expansion is truncated to third order, one finds the usual expression: 
\begin{equation}
\xi_0 = l_3\sqrt{\epsilon}\,.
\label{eq_fit0_xi}
\end{equation}
As seen in fig.~\ref{fig11}, eq.~(\ref{eq_fit0_xi}) accounts for the experimental data over a rather restricted range of accelerations and the $\xi_0^5$ term should be included~\cite{Wernet:2001}. In this case, solving eq.~(\ref{eq.rest}) leads to: 
\begin{equation}
\xi_0^2 = \frac{l_5^4}{2 l_3^2}\left(\sqrt{1+4\epsilon \frac{l_3^4}{ l_5^4}} -1\right)\,, \label{eq_fit_xi}
\end{equation}
which yields a much better description of the $\xi_0$ measurements. 

Once $\xi_0$ (hence $\gamma_0$) is known, the birefringence intensity $I(x,z,t)$ is measured from images obtained between crossed polarizers at $\theta=0$ so that the intensity at the bright spots is maximized. Figure~\ref{fig12}(b) shows $I(x_0,z_0,t)$ recorded at the position of a bright spot. Contrary to the prediction of eq.~(\ref{eqI2}), $I(x_0,z_0,t)$ is not perfectly harmonic but contains a significant subharmonic component. This is most probably due to the finite amplitude of the surface waves which induce a small displacement of the position of the birefringence spots at frequency $f/2$. To decrease the noise in the determination of $I_0$, $I(x_0,z_0,t)$ is averaged over two horizontally adjacent bright spots and $I_0$ is defined as the harmonic component of this mean intensity.

\begin{figure}
\begin{center}
\includegraphics*{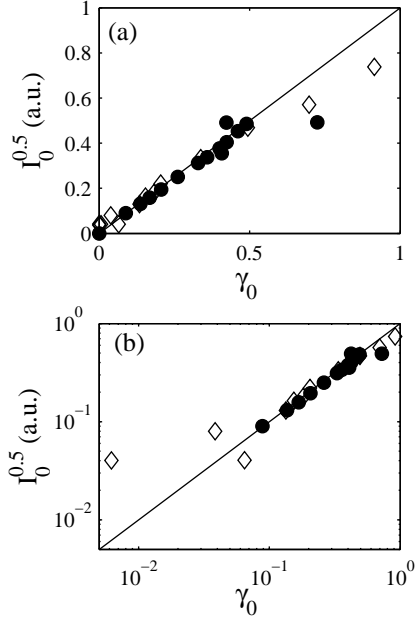}
\end{center}
\caption{Square root of the transmitted intensity $I_0$ versus the deformation $\gamma_0$ for a $5\%$~wt. CPyCl--NaSal solution vibrated at $f=65$~Hz ($\diamond$) and for a $8\%$~wt. CPyCl--NaSal solution vibrated at $f=80$~Hz ($\bullet$). The continuous line is the first diagonal. (a) Linear scales. (b) Logarithmic scales. }
\label{fig13}
\end{figure}

Figure~\ref{fig13} presents the evolution of $I_0$ as a function of $\gamma_0$ for two different concentrations and driving frequencies. We find that $I_0\propto \gamma_0^2$ up to $a=1.1 a_c$. Beyond this point the surface becomes chaotic. Our conclusion is that the stress-optical rule and eq.~(\ref{eqI2}) hold and that the micellar network does not break, at least for $a<1.1 a_c$. In particular micelles remain entangled close to the instability threshold and nonlinear rheology cannot be invoked to explain deviations from linear stability analysis. Thus, even if the amplitude of the equivalent shear rate falls into the nonlinear regime, it appears that the micellar network remains intact under vibrations. We suggest that this is linked to the short period of the forcing ($1/f\simeq 0.02$~s) compared to the characteristic time of our fluid ($\tau_1 \sim 0.5$--1~s). Rather than the amplitude of the equivalent shear rate, it would probably be more relevant to compare the deformation $\xi_0 k_c$ to the critical deformation $\tau_1 \dot{\gamma}_c$ for disentanglement and alignment. A good approximation is $\tau_1 \dot\gamma_c\simeq 2.6$ while $\xi_0 k_c\simeq 0.2$ for $a=1.1 a_c$. Since $\xi_0 k_c\ll\tau_1 \dot\gamma_c$, the fluid deformation over one period of the forcing is too small to break the network and induce the alignment transition.  

\section*{Conclusion}

In this paper we have presented new results for the Faraday instability in wormlike micelles. We have shown that the strong viscoelasticity of these solutions gives rise to original features characterized by oscillations of the critical acceleration and wave number. We clearly linked these oscillations to the presence of standing elastic waves across the cell. Velocimetry experiments have provided evidence that such elastic waves generate counterrotating rolls so that the velocity field does not decrease exponentially with depth. Linear stability analysis was shown to predict and fit both these phenomena. Finally birefringence measurements have shown that the deformation generated at the interface tends to orient the micellar network but is not large enough to induce disentanglement.

We believe that the present experiments may contain worthwile implications for rheological studies of strongly viscoelastic fluids. As noted above and as already mentioned in ref.~\cite{Raynal:1999}, the Faraday instability constitutes an easy way to probe a complex fluid in an oscillating elongational flow at frequencies above 30~Hz. One may thus wonder whether rheological information can be gained from measurements under vertical vibrations and how the Faraday experiment may be used as a ``high'' frequency rheometer. For instance, we tried to infer the rheological properties of our micellar solutions by fitting our velocity field measurements to equations~(\ref{eqv}--\ref{eqq}) without assuming any rheological model \cite{BallestaPhD:2006}. Although the results did not prove conclusive enough (probably due to boundary effects as mentioned at the end of sect.~\ref{sasl}), this may still constitute an interesting direction for future research. Finally another perspective for this work would be to study complex fluids with shorter characteristic times, {\it e.g.} liquid crystals, in which the surface waves may induce microstructural changes.

The authors wish to thank P.~Lettinga and B.~Pouligny for fruitful discussions. S.~Lerouge is thanked for advice on the birefringence experiments and P.~Snabre for technical help on the PIV measurements.

\end{document}